\documentclass{book}    
\usepackage{piers}  
\begin{document}

\title{Validity of Image Theorems under Spherical Geometry}
\maketitle

\author      {Shaolin Liao}
\affiliation {Argonne National Laboratory}
\address     {}
\city        {Argonne, IL}
\postalcode  {}
\country     {USA}
\email       {sliao@anl.gov}  
\misc        { }  
\nomakeauthor

\author      {Sasan Bakhtiari}
\affiliation {Argonne National Laboratory}
\address     {}
\city        {Argonne, IL}
\postalcode  {}
\country     {USA}
\email       {bakhtiari@anl.gov}  
\misc        { }  
\nomakeauthor
\author      {Henry Soekmadji}
\affiliation {Hamilton Sundstrand}
\address     {}
\city        {Rockford, IL}
\postalcode  {}
\country     {USA}
\email       {henry.soekmadji@hs.utc.com}  
\misc        { }  
\nomakeauthor

\begin{authors}

{\bf Shaolin Liao}$^{1}$, {\bf Sasan Bakhtiari}$^{1}$, {\bf and Henry Soekmadji}$^{2}$\\
\medskip
$^{1}$Argonne National Laboratory, USA\\
$^{2}$Hamilton Sundstrand, USA

\end{authors}

\begin{paper}

\begin{piersabstract}
This paper deals with different image theorems, i.e., Love's equivalence principle,   the induction equivalence principle and the physical optics equivalence principle, in the spherical geometry.  The deviation of image theorem approximation is quantified by comparing the  modal expansion coefficients between the electromagnetic field obtained from the image approximation and the exact electromagnetic field for the spherical geometry. Two different methods, i.e., the vector potential method through the spherical addition theorem and the dyadic Green's function method, are used to do the analysis. Applications of the spherical imaging theorems include metal mirror design and other electrically-large object scattering.
\end{piersabstract}

\psection{Introduction}
Different image theorems have been widely used for electromagnetic modeling of mirrors and lens antenna \cite{cylindrical}-\cite{Perkins}. In  \cite{Perkins},   Rong and Perkins applied the image theorems   to   mirror system design   for   high-power   gyrotrons. The author also theoretically evaluate the validity of the image theorems in the cylindrical geometry \cite{cylindrical}.  In this article, following similar procedures in \cite{cylindrical}, a closed-form formula for the discrepancy parameter, which is defined as the ratio of the spherical modal coefficient for image theorem to that of the exact field,  has been derived for the spherical geometry.

\psection{Image Theorems in the Spherical   Geometry}

Fig. \ref{IEEE_sphere} shows the spherical geometry for image theorem analysis.

 \begin{figure}[t] 
 \centering 
\includegraphics[scale=0.7]{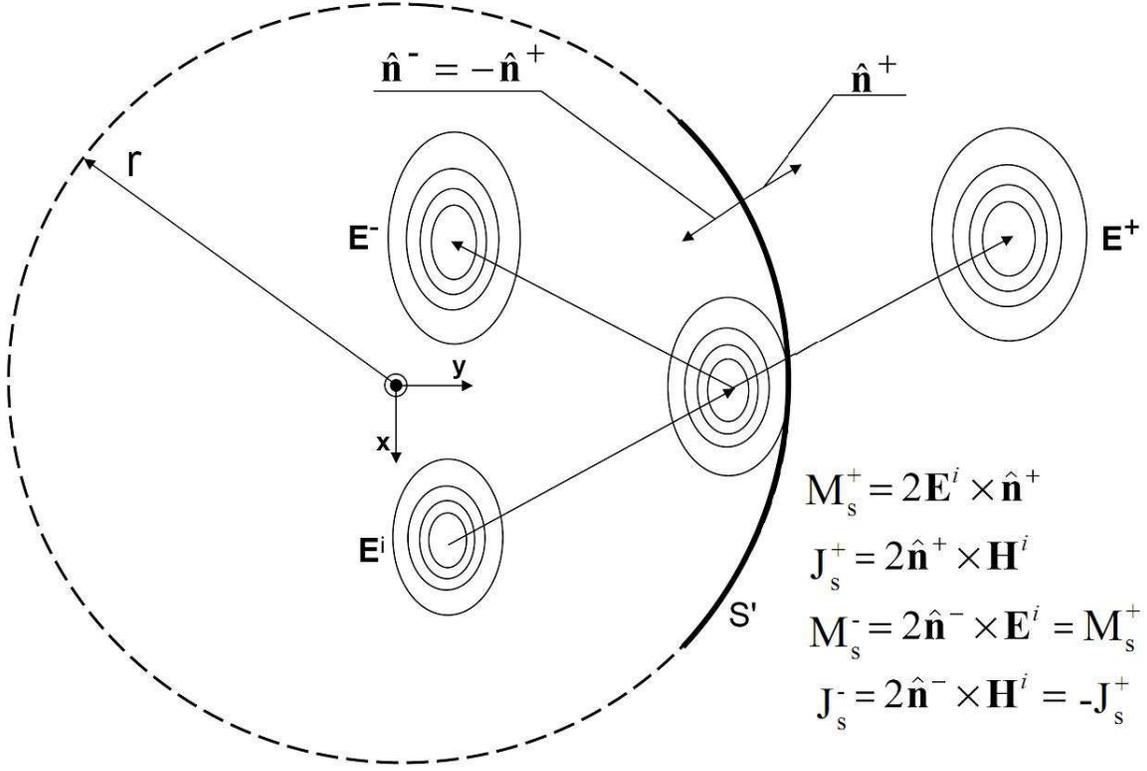}
\caption{Image theorem in the spherical geometry:  the incident field ${\bf
E}^i$ propagates onto spherical surface $S'$, then it may
forward-propagate to ${\bf E}^+$ or it could be back-scattered to
${\bf E}^-$, depending on whether surface $S'$ as a fictitious surface where  the equivalence theorem
 applies on a PEC surface.  $\hat{\bf n}^+$ and  $\hat{\bf n}^-$    are the outward and
 inward surface normals on spherical surface $S'$ respectively.
 ${\bf M}_s$ and ${\bf J}_s$ are equivalent surface currents for
 Love's equivalence theorem. ${\bf M}_s^+$ is the image
 approximation of Love's theorem and ${\bf M}_s^-$ is the image approximation for the
 induction theorem.
 }
 \label{IEEE_sphere}
\end{figure}

\paragraph{  The Vector Potential Method}

\subparagraph{The spherical modal expansion}

In  spherical coordinates, the electrical vector potential  $  { \bf F} ({\bf
r})$ for ${ \bf M}_s  {\bf (r')}$   is given as \cite{Harrington},
\cite{Stratton},
\begin{eqnarray}\label{FM_sph}
    { \bf F}  ({\bf r})  = \epsilon_0  \int \! \! \int_{S'} dS' \    { \bf M}_s  {\bf (r')}
g({\bf r- r' }) = \frac{-j k  \epsilon_0}{4 \pi}  \int \! \! \int_{S'} dS' \     { \bf M}_s {\bf
(r')}    h_0^{(2)}\left(k[r-r'] \right)
 \end{eqnarray}
where, $h_0^{(2)}$ is    spherical Hankel function
of the second kind of order 0.  According to the    spherical addition
theorem   \cite{Harrington}, \cite{Stratton},


\begin{eqnarray}\label{Addition_sph}
h_0^{(2)}\left(k[r-r'] \right)   =
\sum_{n=0}^\infty (2n+1)  j_n( k r')  h_n^{(2)} ( k r) \\
\times \sum_{m=0}^n (2-\delta_m^0) \frac{(n-m)!}{(n+m)!} P_n^m
(\theta') P_n^m (\theta) \cos m(\phi-\phi')  \nonumber
\end{eqnarray}
where, $j_n $ is the spherical Bessel function of
the first kind of integral order n; $P_n^m$ is the associated Legendre polynomial  and  $\delta_m^0$ is the
Kronecker delta function ($\delta_m^0=1$ for m=0 and $\delta_m^0=0$
for m$\neq 0$). Substituting (\ref{Addition_sph}) into
(\ref{FM_sph}), the modal expansion of ${ { \bf F}  (r)}$ is
obtained as,


\begin{eqnarray}\label{ModalF_sph}
  { \bf F} ({\bf r}) & =  &  \sum_{n=0}^\infty  \sum_{m=0}^n { \bf f}^{\hbox{\tiny{\bf M}$_s$}}_{\hbox{\tiny TE}} (n,m)  h_n^{(2)} ( k r)  P_n^m
  (\theta) \begin{array}{c} \cos m\phi \\ \sin m\phi  \end{array} \nonumber \\
    { \bf f}^{m,\hbox{\tiny{\bf M}$_s$} }_{n,\hbox{\tiny \ TE}} & = & \chi \int \! \! \int_{S'} dS' \     { \bf M}_s {\bf
(r')}  j_n( k r')  P_n^m (\theta') \begin{array}{c} \cos m\phi' \\
\sin m\phi'
\end{array}      \nonumber
\\
 \chi  & = &  (2-\delta_m^0) \frac{-j k \epsilon_0}{4 \pi}
 \frac{(2n+1)(n-m)!}{(n+m)!}  \ .
\end{eqnarray}


The near field to far field transform of (\ref{ModalF_sph}) in the spherical
coordinate is given as \cite{Yaghjian},


\begin{eqnarray}\label{FF_sph}
 { \bf F} ( {\bf r})\vline_{r \rightarrow \infty}   =   \frac{j e^{-j k r}}{kr}  \sum_{n=0}^\infty  \sum_{m=0}^n  j^n  { \bf f}^{\hbox{\tiny{\bf M}$_s$}}_{\hbox{\tiny TE}} (n,m)   P_n^m
  (\theta) \begin{array}{c} \cos m\phi \\ \sin m\phi  \end{array}
\end{eqnarray}

The duality relation can be used
to obtain the magnetic vector potential ${ \bf A} ({\bf r})$ for the ${\bf J}_s$ approximation as follows,

\begin{eqnarray}
  { \bf A} ({\bf r}) & =  &  \sum_{n=0}^\infty  \sum_{m=0}^n { \bf g}^{\hbox{\tiny{\bf M}$_s$}}_{\hbox{\tiny TE}} (n,m)  h_n^{(2)} ( k r)  P_n^m
  (\theta) \begin{array}{c} \cos m\phi \\ \sin m\phi  \end{array} \nonumber \\
    { \bf g}^{m,\hbox{\tiny{\bf M}$_s$} }_{n,\hbox{\tiny \ TE}} & = & \chi'  \int \! \! \int_{S'} dS' \     {\bf J}_s {\bf
(r')}  j_n( k r')  P_n^m (\theta') \begin{array}{c} \cos m\phi' \\
\sin m\phi'
\end{array}      \nonumber
\\
 \chi'  & = &  (2-\delta_m^0) \frac{-j k \mu_0}{4 \pi}
 \frac{(2n+1)(n-m)!}{(n+m)!}  \ .
\end{eqnarray}

\subparagraph{The  back-scattered and forward-propagating waves}

 Similar to the cylindrical geometry, we can separate (\ref{ModalF_sph})
 into   back-scattered and forward-propagating waves as,
\begin{eqnarray}\label{Twoparts_sph}
j_n(k r')  =  \frac{1}{2} \left\{ h_n^{(1)}(k r')
+ h_n^{(2)}(k r') \right\}
\end{eqnarray}
\begin{eqnarray}
 { \bf f}^{m,\hbox{\tiny{\bf M}$_s$} \pm}_{n,\hbox{\tiny \ TE}}   =
\frac{\chi}{2}  \int \! \! \int_{S'} dS' \     { \bf M}_s {\bf
(r')}  h_n^{(1), (2)} ( k r')  P_n^m (\theta') \begin{array}{c} \cos m\phi' \\
\sin m\phi'
\end{array}
\nonumber
\end{eqnarray}

Since the spherical harmonics is a complete basis set, we can always  express the initial incident electric field  $ {\bf E}( { \bf r'} )
$  on the initial spherical surface $S'$  with radius of $r_0$  (in
Figure \ref{IEEE_sphere}) as follows,
\begin{eqnarray}\label{Incident_sph}
  { \bf E} ( {\bf r_0}) & = &    \sum_{n=0}^\infty  \sum_{m=0}^n  a_{n,o}^{m,e}  {\bf M}^{m,e+}_{n,o} ( {\bf r_0}) + b_{n,o}^{m,e}  {\bf N}^{m,e+}_{n,o}
  ({\bf r_0})  \nonumber
\end{eqnarray}
\begin{eqnarray} \label{TETM_sph}
  \psi^{m,e+}_{n,o} ({\bf r_0})&=& h_m^{(2)}( k r_0)
P_n^m(\cos \theta')    \begin{array}{c} \cos(m \phi') \\ \sin(m
\phi')
\end{array}    \nonumber \\
  {\bf L}^{m,e+}_{n,o} ({\bf r_0})  &= &  \nabla   \psi^{m,e+}_{n,o} ({\bf r_0})   \nonumber \\
 {\bf M}^{m,e+}_{n,o} ({\bf r_0}) & = & \nabla \times \left\{ {\bf a}_r  r  \psi^{m,e+}_{n,o} ({\bf r_0}) \right\} \nonumber \\
  {\bf N}^{m,e+}_{n,o} ({\bf r_0}) & = & \frac{1}{k} \nabla \times  {\bf M}^{m,e+}_{n,o} ({\bf
  r_0}) \ .
\end{eqnarray}

From (\ref{ModalF_sph}) and noting that $ {\bf M}_s^+({\bf r_0})= 2 {
\bf E} ({\bf r_0}) \times {\bf a}_r $, on  spherical surface $S'$ in
Figure \ref{IEEE_sphere},


\begin{eqnarray}\label{Scatter_sph}
    \tilde{ \bf E} (r_0)     =  -\frac{1}{ \epsilon_0}
\sum_{n=0}^\infty  \sum_{m=0}^n    \left\{ {\bf L}^{m,e+}_{n,o}
({\bf r}) \times   { \bf f}^{m,\hbox{\tiny{\bf M}$_s$}
}_{n,\hbox{\tiny \ TE}}
 \right\}
\end{eqnarray}
\begin{eqnarray}
  {\bf L}^{m,e+}_{n,o} ({\bf r_0})  &= &  \nabla   \psi^{m,e+}_{n,o} ({\bf r_0})   \nonumber
\end{eqnarray}

The approximate field   $ \tilde{ \bf E} ({\bf r}_0)$  on the
initial spherical surface  $S'$   is obtained from
(\ref{ModalF_sph}) through  image theorem approximation,
\begin{eqnarray}\label{zeta_sph}
\tilde{ \bf E} ({\bf r}_0)   =  \sum_{n=0}^\infty \sum_{m=0}^n
\tilde{a}_{n,o}^{m,e} {\bf M}^{m,e}_{n,o}  ( {\bf r}_0)   +
\tilde{b}_{n,o}^{m,e} {\bf N}^{m,e+}_{n,o}
  ({\bf
r}_0)
\end{eqnarray}

Now the deviation of the spherical coefficients $\tilde{a}_{n,o}^{m,e},  \tilde{b}_{n,o}^{m,e}$ in Eq. (\ref{zeta_sph}) from their exact values ${a}_{n,o}^{m,e},  {b}_{n,o}^{m,e}$ in Eq. (\ref{Incident_sph}) is defined as the discrepancy parameters $\zeta$,
\begin{eqnarray}
\zeta_{\hbox{\tiny TE}}^{\hbox{\tiny \bf M}_s}  =
\frac{\tilde{a}_{n,o}^{m,e}}{ {a}_{n,o}^{m,e}} = -j 2 k r_0
h_n^{(2)}(kr_0) \frac{\partial [ kr
j_n(kr)] }{\partial kr}\vline_{\ r=r_0}  \nonumber  \\
\zeta_{\hbox{\tiny TM}}^{\hbox{\tiny \bf M}_s}  =
\frac{\tilde{b}_{n,o}^{m,e}}{ {b}_{n,o}^{m,e}} = j 2 k r_0 j_n(kr_0)
\frac{\partial [ kr h_n^{(2)}(kr)  ] }{\partial kr}\vline_{\ r=r_0}
\nonumber
\end{eqnarray}
and,
\begin{eqnarray}\label{zetapm_sph}
\zeta_{\hbox{\tiny TE}}^{\hbox{\tiny \bf M}_s,\pm} =  -j 2 k r_0
h_n^{(2)}(kr_0) \frac{\partial [ kr h_n^{(1),(2)}(kr)] }{\partial
kr}\vline_{\ r=r_0}  \\
\zeta_{\hbox{\tiny TM}}^{\hbox{\tiny \bf M}_s,\pm}   =   j 2 k r_0
h_n^{(1),(2)}(kr_0) \frac{\partial [ kr h_n^{(2)}(kr)] }{\partial
kr}\vline_{\ r=r_0}  \ . \nonumber
\end{eqnarray}


Similar expressions exist for ${\bf J}_s$ image approximation,
\begin{eqnarray}\label{J_sph}
\zeta_{\hbox{\tiny TE}}^{\hbox{\tiny \bf J}_s}  = \zeta_{\hbox{\tiny
TM}}^{\hbox{\tiny \bf M}_s} , \ \ \zeta_{\hbox{\tiny
TM}}^{\hbox{\tiny \bf J}_s} = \zeta_{\hbox{\tiny TE}}^{\hbox{\tiny
\bf M}_s}
\end{eqnarray}
\begin{eqnarray}
\zeta_{\hbox{\tiny TE}}^{\hbox{\tiny \bf M}_s,\pm}
=\zeta_{\hbox{\tiny TM}}^{\hbox{\tiny \bf J}_s,\pm}    =
[\zeta_{\hbox{\tiny TE}}^{\hbox{\tiny \bf J}_s,\pm}]^\ast  =
[\zeta_{\hbox{\tiny TM}}^{\hbox{\tiny \bf M}_s,\pm}]^\ast  \ .
\nonumber
\end{eqnarray}

\paragraph{The Dyadic Green's Function Method}

\  \ The magnetic dyadic Green's function in the spherical coordinate is,
\begin{eqnarray}\label{Dyadic_sph}
\bar{\bf G}_m ({\bf r}, {\bf r'}) & = &  - \frac{{\bf a}_r {\bf
a}_r}{ k^2} \delta( {\bf r - r'} ) - \sum_{n=-\infty}^\infty
  \frac{ j \pi }{ 2 k n (n+1)   }
   \nonumber
\end{eqnarray}
\begin{eqnarray}
  \times   \sum_{m=0}^n  \frac{1}{Q_{nm}} \left\{
   {\bf M}^{m,e}_{n,o} ( {\bf r'}) {\bf M}^{m,e+}_{n,o} ( {\bf r}) + {\bf N}^{m,e}_{n,o}  (  {\bf r'} )
{\bf N}^{m,e+}_{n,o} ( {\bf r})  \right\}   \nonumber \\
 \hbox{and,} \ \  \ \ \ \ \ \   \ \ \ \ \ \     Q_{nm}  =   \frac{2 \pi^2 (n+m)!}{ (2-\delta_m^0) (2n+1)
 (n-m)!}   \ \ \ \ \ \     \ \ \ \
\end{eqnarray}
where ${\bf M}^{m,e}_{n,o} $ (${\bf
N}^{m,e}_{n,o}$) is obtained by replacing $h_n^{(2)}$ with $j_n$ in
${\bf M}^{m,e+}_{n,o}$ (${\bf N}^{m,e+}_{n,o}$).
 The approximate field $\tilde{\bf E} ({\bf r})$  for ${\bf M}_s^+({\bf r'})$ is given as,


\begin{eqnarray}\label{Scatter_dyadic_sph}
 \tilde{\bf E} ({\bf r})   & = &     - \nabla  \times  \int \!\! \int_{S'} dS' \ {\bf M}_s^+({\bf r'}) {\bf
 .} \bar{\bf G}_m ({\bf r}, {\bf r'})
\end{eqnarray}
Substituting (\ref{Dyadic_sph})  into   (\ref{Scatter_dyadic_sph})
and using the orthogonal properties of   spherical modal functions,
the approximate field $ \tilde{\bf E} ( r_0 ) $ on initial spherical
surface $S'$ is obtained as,


\begin{eqnarray}\label{Scatter_dyadic_sph_1}
 \tilde{\bf E} ( r_0 )    =
 \sum_{n=-\infty}^{\infty}  \sum_{m=0}^{n}    \frac{j \pi}{n (n+1) Q_{nm}}
    \begin{array}{c}  {c}_{n,o}^{m,e}  {\bf M}_{n,o}^{m,e+} ({\bf r})  \\
  {d}_{n,o}^{m,e}  {\bf N}_{n,o}^{m,e+} ({\bf r})
\end{array}    \\
\times     \int \! \! \int_{S'} dS' \
   \begin{array}{c}  \hbox{$[{\bf N}_{n,o}^{m,e} ({\bf r'})]^\ast$}   \times  {\bf M}_{n,o}^{m,e+} ({\bf r'})
    \\   \hbox{$[{\bf M}_{n,o}^{m,e} ({\bf r'})]^\ast$}    \times  {\bf N}_{n,o}^{m,e+} ({\bf r'})  \end{array}  \   {\bf    \cdot} \  {\bf
a}_{r'} \ .
    \nonumber
 \end{eqnarray}
 The evaluation of (\ref{Scatter_dyadic_sph_1}) also leads to
 (\ref{zeta_sph}) and (\ref{zetapm_sph}).

\begin{table*}
\caption{Summary of   $\zeta_{\hbox{\tiny TE, TM}}^{+,-} ({\bf m}_s ,
{\bf j}_s) $ for the spherical geometry}
\label{xi_zeta}
\begin{tabular}{cccc}
  \\
  TE/TM modes and  ${\bf M}_s/{\bf J}_s$  & The
relations & $\zeta_{\hbox{\tiny TE, TM}}^{+,-} ({\bf
M}_s ,  {\bf J}_s) $  &      $ r_0 \rightarrow \infty$ \\
  & &  & \\
\hline     \hline  \\
 TE   \& ${\bf M}_s$   /  TM \& ${\bf J}_s$    & &   &    \\
    \vspace*{0.05in}  Sphere: back-scattered wave &   $\zeta_{\hbox{\tiny TE}}^-( {\bf M}_s )=\zeta_{\hbox{\tiny TM}}^-( {\bf J}_s )$ &  $ -j   k r_0 h_n^{(2)}(kr_0) \frac{\partial [ kr
h_n^{(2)}(kr)] }{\partial kr}\vline_{\ r=r_0}$  &  $  (-1)^{n} e^{-j 2 k r_0}$ \\
  \vspace*{0.05in}  Sphere: forward-propagating wave &   $\zeta_{\hbox{\tiny TE}}^+( {\bf M}_s ) = \zeta_{\hbox{\tiny TM}}^+( {\bf J}_s )$ &  $ -j   k r_0 h_n^{(2)}(kr_0) \frac{\partial [ kr
h_n^{(1)}(kr)] }{\partial kr}\vline_{\ r=r_0} $  & 1 \\
   & &  & \\
   TM   \& ${\bf M}_s$    /  TE \& ${\bf J}_s$     & &   & \\
  \vspace*{0.05in}  Sphere: back-scattered wave  &
$\zeta_{\hbox{\tiny TM}}^-( {\bf M}_s )=\zeta_{\hbox{\tiny TE}}^-(
{\bf J}_s )$ &  $ j   k r_0 h_n^{(2)}(kr_0)  \frac{\partial [ kr
h_n^{(2)}(kr) ] }{\partial kr}\vline_{\ r=r_0}$  & $ - (-1)^{n} e^{-j 2 k r_0}$ \\
 \vspace*{0.05in}  Sphere: forward-propagating wave &   $\zeta_{\hbox{\tiny TM}}^+( {\bf M}_s ) = \zeta_{\hbox{\tiny TE}}^+( {\bf J}_s )$ & $ [\zeta_{\hbox{\tiny TE}}^+( {\bf M}_s )]^\ast / [\zeta_{\hbox{\tiny TM}}^+( {\bf J}_s )]^\ast$ & 1 \\
   & & & \\
   \hline    \hline   \\
\end{tabular} 
\end{table*}

\paragraph{The Analytical Formula for   Image Theorems   in the Spherical  Geometry}

Similar to the cylindrical geometry,  $\zeta_{\hbox{\tiny
TE,TM}}^{\hbox{\tiny \bf M}_s,\hbox{\tiny \bf J}_s +} $ in
(\ref{zetapm_sph}) and (\ref{J_sph}) can be considered as
theoretical formulas for evaluation  of the image theorems for
narrow-band fields in the spherical geometry. The large argument
asymptotic behaviors of $\zeta_{\hbox{\tiny TE,TM}}^{\hbox{\tiny \bf
M}_s,\hbox{\tiny \bf J}_s +} $  for $r_0 \rightarrow \infty$
  can be obtained by noting that,
\begin{eqnarray}
 h_n^{(2)}(k r_0) =  [h_n^{(1)}(k r_0)]^\ast   \sim
 \frac{1}{k r_0} j^{(n+1)} e^{-j k r_0}, \
 k r_0 \rightarrow \infty \nonumber
 \end{eqnarray}
\begin{eqnarray}
\zeta_{\hbox{\tiny TE,TM}}^{\hbox{\tiny \bf M}_s,\hbox{\tiny \bf
J}_s +} \vline_{ \ r_0 \rightarrow \infty}      =   1  \ .
\end{eqnarray}

\begin{figure}[t]
\centering 
\includegraphics[scale=0.8]{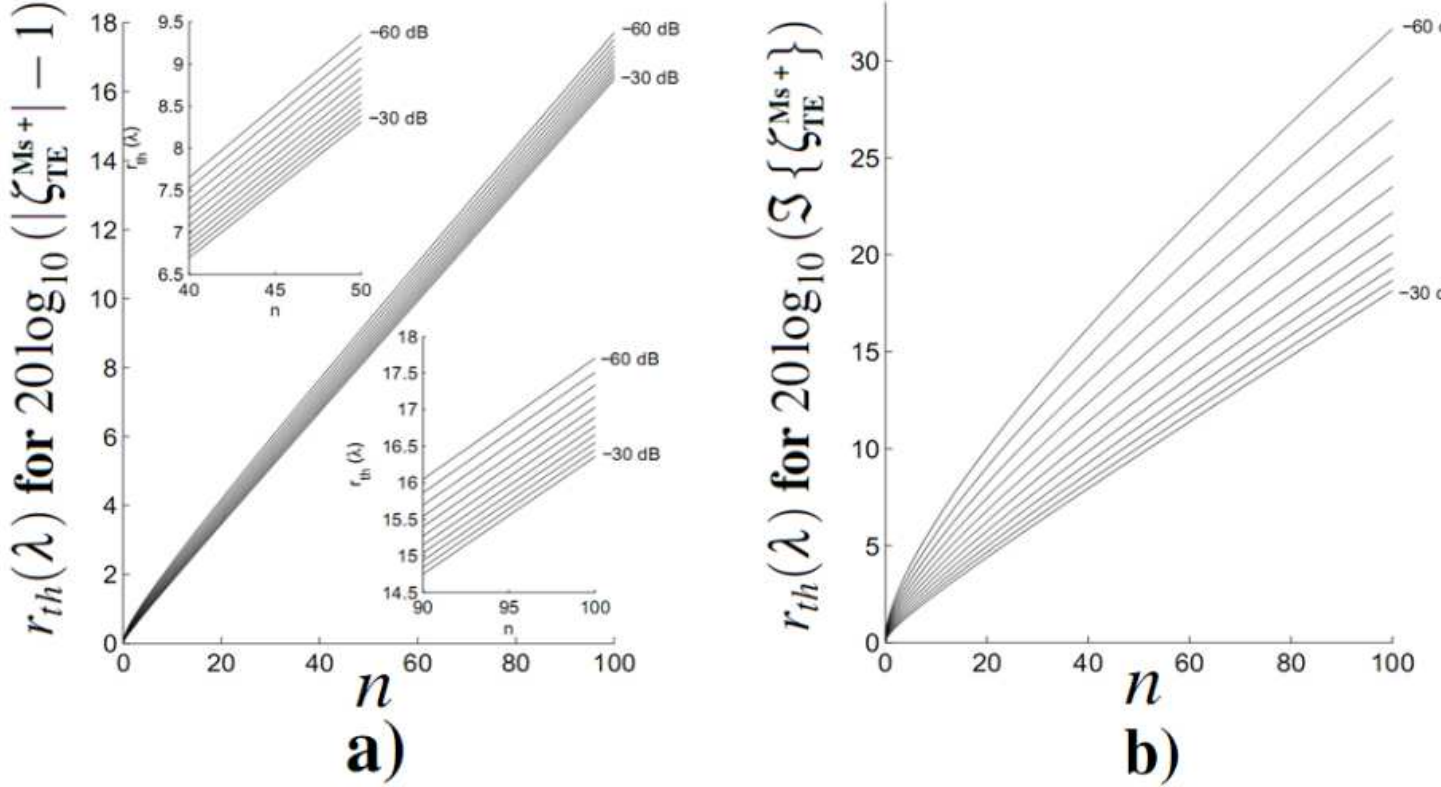} 
 \caption{ The spherical geometry - threshold radii $r_{\hbox{\tiny th}}$ Vs.
n=0 to 100, for different accuracies, from $-60$ dB to $-30$ dB (in
$10$ dB increment, from bottom to top): a)
 the magnitudes $20 \log_{10}(|\zeta_{\hbox{\tiny TE
}}^{\hbox{\tiny \bf M}_s +}|-1)$, and b) the imaginary parts  $20
\log_{10}[\Im(\zeta_{\hbox{\tiny TE }}^{\hbox{\tiny \bf M}_s +})]$.
The inset plots in a) are used to make the display clearer. Similar
to the cylindrical geometry,  imaginary parts  $\zeta_{\hbox{\tiny
TE }}^{\hbox{\tiny \bf M}_s +}$ require larger threshold radii
$r_{th} $ for the same accuracy.
  }
 \label{zeta_rho_sph}
\end{figure}

\psection{Results and Discussion}

TABLE \ref{xi_zeta} summarizes the properties of
$\zeta_{\hbox{\tiny TE,TM }}^{\hbox{\tiny \bf M}_s,\hbox{\tiny \bf
J}_s \pm}$, for the back-scattered and forward-propagating waves
respectively. For $r_0 \rightarrow \infty$,
$\zeta_{\hbox{\tiny TE,TM }}^{\hbox{\tiny \bf M}_s,\hbox{\tiny \bf
J}_s  } = \zeta_{\hbox{\tiny TE,TM }}^{\hbox{\tiny \bf
M}_s,\hbox{\tiny \bf J}_s +} + \zeta_{\hbox{\tiny TE,TM
}}^{\hbox{\tiny \bf M}_s,\hbox{\tiny \bf J}_s -}$ shows fast
oscillations, which can be seen from TABLE \ref{xi_zeta}. Mathematically, the
oscillations only appear as  modal expansion coefficients and
disappear after the implementation of  the double sums in (\ref{zeta_sph}).
Physically, the oscillations are due to back-scattered fields, which
approach 0 for $r_0 \rightarrow \infty$. For example,
consider $\zeta_{\hbox{\tiny TE  }}^{\hbox{\tiny \bf M}_s -}$ in
(\ref{zetapm_sph}),


\begin{eqnarray}\label{Eminus}
\tilde{ \bf E}^- (r_0, \phi)   =  \sum_{n=0}^\infty \sum_{m=0}^n
\left\{ \zeta_{\hbox{\tiny TE  }}^{\hbox{\tiny \bf M}_s -}  \
{c}_{n,o}^{m,e} {\bf M}^{m,e+}_{n,o}  ( {\bf r}_0)
  +   \zeta_{\hbox{\tiny TE  }}^{\hbox{\tiny \bf M}_s -} \ {b}_{n,o}^{m,e} {\bf
N}^{m,e+}_{n,o}   ({\bf r}_0)  \right\}
\end{eqnarray}


Changing the variable $\phi' = \phi - \pi$ and letting  $r_0
\rightarrow \infty$, from TABLE \ref{xi_zeta}, (\ref{Eminus})
reduces to,


\begin{eqnarray}
\tilde{ \bf E}^- (r_0, \phi')\vline_{ \ r_0 \rightarrow \infty} =
\sum_{n=0}^\infty \sum_{m=0}^n e^{-j 2 k r_0}  \left\{
{c}_{n,o}^{m,e} {\bf M}^{m,e+}_{n,o}  ( {\bf r}_0)
  -    {b}_{n,o}^{m,e} {\bf
N}^{m,e+}_{n,o}   ({\bf r}_0)  \right\}.
\end{eqnarray}


 Now, the back-scattered field   $\tilde{ \bf E}^- (r_0, \phi')\vline_{ \ r_0
\rightarrow \infty} \rightarrow 0$  due to the fast variation phase
term $e^{-j 2 k r_0}$, which  means that the oscillation in
$\zeta_{\hbox{\tiny TE  }}^{\hbox{\tiny \bf M}_s -}$ doesn't appear
in the actual field  evaluation for  $r_0 \rightarrow \infty$ .

 Based on the above discussion, $\zeta_{\hbox{\tiny TE,TM }}^{\hbox{\tiny \bf M}_s,\hbox{\tiny \bf J}_s \pm}$ is the theoretical formula of interest to evaluate the
validity of image theorems. 

  It is also helpful to plot the corresponding
threshold radius $r_{\hbox{\tiny th}}$ with respect to n, for both
$20 \log_{10}(|\zeta_{\hbox{\tiny TE }}^{\hbox{\tiny \bf M}_s
+}|-1)$ and $20 \log_{10}\{\Im[\zeta_{\hbox{\tiny TE }}^{\hbox{\tiny
\bf M}_s  +}]\}$, with   different accuracies ranging from $-60$ dB
to $-30$ dB (in $3$ dB increment), as in Fig. \ref{zeta_rho_sph}.
It can be seen from  Fig.
\ref{zeta_rho_sph} that, in order to achieve an accuracy of $-30$ dB
for $|\zeta_{\hbox{\tiny TE }}^{\hbox{\tiny \bf M}_s  +}|$ (with
respect to 1), $r_{\hbox{\tiny th}} \sim 8 \lambda$    and
  $r_{\hbox{\tiny th}} \sim 16 \lambda$   for $n=50$ and $n=100$ respectively.
However, for the imaginary part $\Im[\zeta_{\hbox{\tiny TE
}}^{\hbox{\tiny \bf M}_s  +}]$, $r_{\hbox{\tiny th}} \sim 9.5
\lambda$   and
  $r_{\hbox{\tiny th}} \sim 18 \lambda$  are required for $n=50$ and $n=100$
respectively, which again implies that the imaginary part
$\zeta_{\hbox{\tiny TE }}^{\hbox{\tiny \bf M}_s  +}$ dominates the
accuracy of image theorems.

\psection{Conclusion}

For spherical geometry,  the theoretical formulas  for  evaluation of the  image theorems
(both ${\bf M}_s$ and ${\bf J}_s$ approximations) have been derived
through two equivalent methods - the vector potential method and the
dyadic Green's function method, for both TE and TM modes.  The ratio of the spherical modal coefficient of the image theorem to that of the exact field is used as the criterion to determine the validity of the image theorem.

\end{paper}


\begin{thebibliography}{99}


\bibitem{cylindrical}
 Shaolin Liao and R. J. Vernon, ``On the Image Approximation for Electromagnetic Wave Propagation and PEC Scattering in Cylindrical Harmonics", Progress In Electromagnetics Research, PIER 66, 65-88, 2006.

\bibitem{liao_beam-shaping_2007} Shaolin Liao, ``Beam-shaping PEC Mirror Phase Corrector Design," {\it PIERS Online}, 3(4):392-396, 2007.
 
\bibitem{Shaolin_30}  S.-L. Liao and R. J.  Vernon,
``A new fast algorithm for field propagation
 between arbitrary smooth surfaces",   { \it the joint  30$^{\hbox{\tiny th}}$ Infrared and
Millimeter Waves and 13$^{\hbox{\tiny th}}$ International Conference
on Terahertz Electronics}, Williamsburg, Virginia, USA, 2005, ISBN:
0-7803-9348-1, INSPEC number: 8788764, DOI:
10.1109/ICIMW.2005.1572687, Vol. 2,  pp.~\mbox{606-607}.


\bibitem{Shaolin_31_1} S.-L. Liao and R. J. Vernon, ``The near-field and   far-field properties of the cylindrical modal
 expansions with  application in the image theorem," { \sl the   31$^{\hbox{\tiny st}}$ Int.
Conf. on Infrared and Millimeter Waves,} Shanghai,
 China, IEEE MTT, Catalog Number: 06EX1385C,
ISBN: 1-4244-0400-2, Sep. 18-22, 2006.



\bibitem{Shaolin_31_2} S.-L. Liao and R. J.  Vernon, ``The cylindrical Taylor-interpolation FFT  algorithm,"
 { \sl the  31$^{\hbox{\tiny st}}$ Int. Conf. on Infrared and Millimeter Waves,} Shanghai,
 China, IEEE MTT, Catalog Number: 06EX1385C,
ISBN: 1-4244-0400-2, Sep. 18-22, 2006.
 
  \bibitem{Shaolin_JEMWA} S.-L. Liao and R. J.  Vernon,  ``Sub-THz beam-shaping mirror  designs  for quasi-optical mode
  converter in high-power  gyrotrons",
  {\it J. Electromagn. Waves and Appl.}, scheduled for volume 21, number 4, page 425-439,
  2007.
  
  
\bibitem{liao_near-field_2006}
Shaolin Liao and R.J. Vernon, ``A new fast algorithm for calculating near-field propagation between arbitrary smooth surfaces," In {\it 2005 Joint 30th International Conference on Infrared and Millimeter Waves and 13th International Conference on Terahertz Electronics}, volume 2, pages 606-607 vol. 2, September 2005. ISSN: 2162-2035.

\bibitem{shaolin_liao_new_2005}
Shaolin Liao, Henry Soekmadji, and Ronald J. Vernon, ``On Fast Computation of Electromagnetic Wave Propagation through FFT," In {\it 2006 7th International Symposium on Antennas Propagation EM Theory}, pages 1-4, October 2006.
  

\bibitem{liao_fast_2007}
Shaolin Liao, ``Fast Computation of Electromagnetic Wave Propagation and Scattering for Quasi-cylindrical Geometry," {\it PIERS Online}, 3(1):96-100, 2007.

\bibitem{liao_validity_2007} Shaolin Liao, ``On the validity of physical optics  for narrow-band
beam scattering and diffraction   from the   open cylindrical
surface," {\it Progress in Electromagnetics Research Symposium (PIERS)}, vol. 3, no. 2, pp. 158–162 Mar., 2007. arXiv:physics/3252668. DOI: 10.2529/PIERS060906142312

\bibitem{liao_high-efficiency_2008} Shaolin Liao, Ronald J. Vernon, and Jeffrey Neilson, ``A high-efficiency four-frequency mode
converter design with small output angle variation for a step-tunable gyrotron," In {\it 2008 33rd International Conference on Infrared, Millimeter and Terahertz Waves}, pages 1-2, September 2008. ISSN: 2162-2035.

\bibitem{liao_four-frequency_2009}
S. Liao, R. J. Vernon, and J. Neilson, ``A four-frequency mode converter with small output angle variation for a step-tunable gyrotron," In {\it Electron Cyclotron Emission and Electron Cyclotron Resonance Heating (EC-15)}, pages 477-482. WORLD SCIENTIFIC, April 2009.

\bibitem{vernon_high-power_2015}
Ronald J. Vernon, ``High-Power Microwave Transmission and Mode Conversion Program," Technical Report DOEUW52122, Univ. of Wisconsin, Madison, WI (United States), August 2015.

\bibitem{liao_multi-frequency_2008}
Shaolin Liao, {\it Multi-frequency beam-shaping mirror system design for high-power gyrotrons: theory, algorithms and methods}, Ph.D. Thesis, University of Wisconsin at Madison, USA, 2008. AAI3314260 ISBN-13: 9780549633167.

\bibitem{liao_fast_2007-1}
Shaolin Liao and Ronald J. Vernon, ``A Fast Algorithm for Wave Propagation from a Plane or a Cylindrical Surface," {\it International Journal of Infrared and Millimeter Waves}, 28(6):479-490, June 2007.
 

\bibitem{liao_miter_2009}
Shaolin Liao, ``Miter Bend Mirror Design for Corrugated Waveguides," {\it Progress In Electromagnetics Research}, 10:157-162, 2009.  

\bibitem{liao_fast_2009}
Shaolin Liao and Ronald J. Vernon, ``A Fast Algorithm for Computation of Electromagnetic Wave Propagation in Half-Space," {\it IEEE Transactions on Antennas and Propagation}, 57(7):2068-2075, July 2009.  

\bibitem{liao_efficient_2011}
Shaolin Liao, N. Gopalsami, A. Venugopal, A. Heifetz, and A. C. Raptis, ``An efficient iterative algorithm for computation of scattering from dielectric objects," {\it Optics Express}, 19(4):3304-3315, February 2011. Publisher: Optical Society of America.

\bibitem{liao_spectral-domain_2019}
Shaolin Liao, ``Spectral-domain MOM for Planar Meta-materials of Arbitrary Aperture Wave-guide Array," In {\it 2019 IEEE MTT-S International Conference on Numerical Electromagnetic and Multiphysics Modeling and Optimization (NEMO)}, pages 1-4, May 2019.

\bibitem{Perkins} Michael P. Perkins and Ronald J.  Vernon,  Iterative design of a cylinder-based beam-shaping
 mirror pair for use in a gyrotron internal quasi-optical mode converter,  the  29$^{\hbox{\tiny th}}$
Int. Conf. on Infrared and Millimeter Waves, Karlsruhe, Germany,
Sep. 27-Oct. 1, 2004.

\bibitem{Harrington}
Roger F. Harrington, {\sl Time-Harmonic Electromagnetic Fields,}
McGraw-Hill, Inc., 1961.

\bibitem{Stratton}
J. A. Stratton, {\sl Electromagnetic Theory,} McGraw-Hill, Inc.,
1941.


\bibitem{Yaghjian}
A. D. Yaghjian,  An overview of near-field antenna measurements,
 IEEE Trans. on Antennas and Propagat., 34(1) (1986) 30-45.




\end{thebibliography}
\end{document}